\documentstyle[aps,prl,epsf,epsfig,]{revtex}
\begin{document}
\draft
\title{
Laser-Induced Entangled Giant Quasi-Molecules in  Optical Lattices }
\author{Bimalendu Deb$^{1,2}$ and Gershon Kurizki$^{1}$}
\address{
$^{1}$Department of Chemical Physics, Weizmann Institute of Science,
76100 Rehovot, Israel\\
$^{2}$  Physical Research Laboratory,
Ahmedabad-380009, India}
\date{\today}
\maketitle
\begin{abstract}
We analyze atom-atom interactions in  optical lattices due to a
laser-induced long-range interatomic force which prevails over
the usual London-van der-Waals forces.   This force, which can be
generated by an intense laser field at a wavelength longer than
that of the lattice-generating laser, is shown to bind pairs of
cold atoms trapped at different lattice sites, and cause their
translational quantum correlations (spatial entanglement).
\end{abstract}

The best-known laser-induced  force acting upon or between atoms
is the "dipole force", resulting from the action of the
{\it gradient} of the laser field, which has been extensively 
used to confine cold atoms in  optical traps \cite{traps}.
This force has also been exploited
in optical lattices, created by intersecting laser beams, which can
trap cold atoms 
in periodic potential wells with typical depth of neV and inter-well 
separations on the order of the laser wavelength \cite{optical lattice}. 
Recently, the mechanical effect of another laser-induced force
has come to the fore, that of dipole-dipole interactions generated by
{\it off-resonant} laser beams with small gradient (nearly plane waves)
\cite{induceddd}. In the {\it near-zone} region of multiple,
mutually-detuned and appropriately oriented laser beams, 
such dipole-dipole interactions yield a gravity-like 
interatomic  $1/r$ attraction \cite{duncan}. In contrast to 
the previously considered {\em near-resonant} dipole-dipole interaction 
\cite{brennen} via {\em real} photon exchange between a ground-state atom and its
excited partner, wherein their cooperative decay plays a role, the 
far-off resonant dipole-dipole interaction arises from {\it virtual} 
(nearly decay-free) photon exchange between two ground-state atoms, 
which acquire small induced dipole-moments \cite{induceddd}.

Here we put forward a new effect of the off-resonant,
laser-induced dipole-dipole interaction: its ability to bind
together pairs of ground-state atoms localized at adjacent sites
of an optical lattice. 
The resulting novel 
``giant'' (submicron-size) quasimolecules will remain bound even
if the lattice potential is switched off. 
An important feature of such quasimolecules
 are their {\em translational}
quantum correlations or entanglement, i.e., nonfactorizability of the two-atom 
spatial wavefunction. The predicted effects should be contrasted with 
entanglement between {\it internal} states of closely separated
atoms residing at the same site of a densely occupied optical 
lattice, via the essentially different near-resonant
dipole-dipole interaction \cite{brennen}.

The retarded
dipole-dipole interaction potential between 
atoms A and B, induced by an off-resonant
circularly-polarized,
plane-wave laser is \cite{induceddd}.
\begin{eqnarray}
V_{AB} = &-& \frac{2\pi k^3\alpha^2I}{c}F_\theta(kr);\ \ F_\theta(kr)= \cos(kr
\cos\theta) \nonumber \\ &\times& \left[\frac{\{\cos(kr) + kr
\sin(kr)\} (1-3 \cos^{2}\theta)}{(kr)^{3}} + \frac{(1 +
\cos^{2}\theta) \cos(kr)}{kr}\right].
\end{eqnarray}
Here $\theta$ is the angle between the interatomic axis and
the wave-vector $k=\omega/c$ of the laser, $I$ is the
laser intensity and $\alpha = \frac{2\omega_{A}{\bf
d}^{2}}{\hbar(\omega_{A}^{2}-\omega^{2})}$ is the atomic dynamic
polarizability, with ${\bf d}$ and
$\omega_{A}$  being the atomic dipole moment and transition frequency,
 respectively, assuming that both atoms ($A,B$) have the
same polarizability. In a 3D optical  lattice, the $V_{AB}$ potential 
in Eq. (1) can be attractive for a particular relative position of two 
tightly-bound atoms whilst repulsive for another. 
Yet the lattice
allows us to 
find a direction of propagation of the additional laser beam such that 
$V_{AB}(kr,\theta)$ is attractive  for two atoms located at adjacent 
lattice sites, $k_L r \simeq \pi$, along certain lattice axes, thus making
the quasibinding effect observable, as illustrated below.

For pair binding 
 at adjacent lattice sites to have a distinct spectral signature,
the
binding potential depth $|V_{AB}(r\simeq\pi/k_L)|$ should {\em 
substantially exceed} two parameters:  
 (a) the {\em linewidth for
the off-resonant absorption} of the additional laser, $\hbar\Gamma S$,
where $\Gamma$ is the resonant
transition linewidth and the
saturation parameter $S=\frac{2\Omega^2}{4\delta^2+\Gamma^2}$,
with $\Omega$ being the laser Rabi frequency and
$\delta=\omega-\omega_{A}$ its detuning,
should satisfy
$S<\!<1$, thus keeping the interacting atoms {\em unexcited}, as
required by the validity of Eq. (1) \cite{induceddd,duncan};
(b) {\em the heating energy} $\hbar\Gamma_{\rm heat}$
by the additional laser given by \cite{note} $\Gamma_{\rm
heat}\sim E_R\Gamma_{\rm Ray}^{(inelas)}/|V_{AB}|$, where $E_R=\hbar^2k^2/2m$ is the
recoil energy of an atom with mass $m$, and the inelastic 
Rayleigh-scattering
rate $\Gamma_{\rm Ray}^{\rm (inelas)}$ satisfies \cite{duncan} 
$\hbar\Gamma_{\rm Ray}^{\rm (inelas)}/|V_{AB}|\sim 
f_{\rm LD}(\vec{k}\cdot\vec{r})
/|F_\theta(kr)|$, $f_{\rm LD}(\vec{k}\cdot\vec{r})$ being the
Lamb-Dicke factor of the system, which reduces to $(kr)^2\ll 1$ in the
near zone, and $F_\theta(kr)$ is given by (1).
Conditions (a) and (b) can be
satisfied in the {\em near-zone regime} of $V_{AB}$, at 
$r\simeq\pi/k_L\ll\pi/k$,
such that $F_\theta(kr)\gg 1$ and 
$f_{\rm LD}(\vec{k}\cdot\vec{r})\ll 1$.
This requires the additional laser to
have a much longer wavelength than the lattice period, $k\ll k_L$.

We proceed with the general theoretical framework that allows one
to treat atom-atom scattering in an optical lattice for any
interatomic  potential $V_{AB}(r)$, which is assumed to be a 
function of the interatomic separation $r = |{\bf r}_{A} - {\bf r}_{B}|$ only.
The Hamiltonian of the two interacting atoms is $H = H_{0}+ V_{AB} $, where
\begin{equation}
H_{0} =  - \frac{\hbar^{2}}{2m_{A}} \nabla_{A}^2 -
\frac{\hbar^{2}}{2m_{B}} \nabla_{B}^2   + U_{A}(r_{A}) +
U_{B}(r_{B}),
\end{equation}
$U_{A(B)}$ being the periodic potential of the lattice acting on atom $A(B)$.

The formation of a bound or quasi-bound pair is revealed by the analysis of
the transition matrix
\begin{equation}
{\mathbf{T}} = V_{AB} ({\bf 1} - G V_{AB})^{-1}
\end{equation}
$G$ being the unperturbed Green function (associated with $H_{0}$). 
Let $D$ be the determinant of the matrix $({\bf 1} - G V_{AB})$.
When ${\rm Re}(D) = 0$, a {\it bound state} (resonance with zero width) 
may be formed if its energy $E_{b}$ exceeds the total band energy of the 
two unperturbed atoms. A resonance may occur if ${\rm Re}(D) = 0$ corresponds
to an energy $E_{r}$ {\it within} the unperturbed total band energy.
Such a resonance implies  the existence of two-atom quasi-bound states in the
lattice, but only if its width (its reciprocal lifetime) is given by
the {\it positive} value \cite{callaway}
\begin{equation}
\Gamma_{r} = 2 \frac{{\mathrm Im}(D)}{\frac{d}{dE} {\mathrm
Re}(D)} .
\end{equation}

Information about the two-atom states
can be inferred from Bragg
scattering spectra, which
depend on  the atomic density of states in the lattice. The
change in the two-atom density of states due to the interatomic
interaction potential $V_{AB}$ is given by
\begin{equation}
\delta\rho = -\frac{1}{\pi N} {\text Im}[ \frac{d}{dE}({\mathrm
ln}D)] ,
\end{equation}
$N$ being the normalization constant. In the vicinity of a genuine
resonance located at $E=E_{r}$ (i.e., if the conditions leading to Eq. (4)
are satisfied), Eq. (5) reduces to the Lorentzian shape \cite{callaway}
\begin{equation}
\delta\rho =
\frac{\Gamma_{r}/2\pi}{(E-E_{r})^{2}+\Gamma_{r}^{2}/4} .
\end{equation}
This expression measures {\it the enhancement} of the two-atom density 
of states near $E_r$, relative to that of a perfect optical lattice (without
atom-atom
interaction),
\begin{equation}
\rho^{(0)}(E) = -\frac{1}{\pi N} {\text Im}[{\mathbf{ Tr}}(G)] .
\end{equation}
Experimentally, however, $\Gamma_r$ will be swamped by the much larger
line broadening due to the $\Gamma_{\rm heat}$ discussed above.

We now consider that the
atoms are located near the minima of the periodic potential wells
of the optical lattice, for which
the tight-binding approximation is valid.
As the basis states, we choose the Wannier
functions $\chi_{n}(r-\vec{R}_{\vec{\mu}})$ 
which have  the  property: $ \int d\vec{r}
\chi_{n}(\vec{r}-\vec{R}_{\vec{\mu}}) H_{0}
\chi_{n}(\vec{r}-\vec{R}_{{\vec\mu}^{\prime}})
 = \epsilon_{n}(\vec{R}_{\vec{\mu}}-\vec{R}_{\vec{\mu}^{\prime}})$, 
$\epsilon_{n}$ being the $n$th band energy eigenvalue
and $\vec{R}_{\vec{\mu}}$ the site
location. 
The symmetrized or antisymmetrized product of the Wannier functions of the
two atoms is denoted by
\begin{equation}
|nm; \vec{\mu} \vec{\mu}^{\prime} \rangle ={\mathcal N}[
\chi_{n}(\vec{r}_{A}-\vec{R}_{\vec{\mu}})
\chi_{m}(\vec{r}_{B}-\vec{R}_{\vec{\mu}^{\prime}}) \pm
\chi_{n}(\vec{r}_{B}-\vec{R}_{\vec{\mu}})
\chi_{m}(\vec{r}_{A}-\vec{R}_{\vec{\mu}^{\prime}})],
\end{equation}
where ${\mathcal N}$ is a normalization constant, and + (-) refers to
bosons (fermions). Here we assume that $\vec{\mu}\neq \vec{\mu}^{\prime}$,
i.e. no lattice-site is occupied by {\em more than one atom}. In the plots
and estimates given below only bosons are considered.

In a simple cubic (SC) lattice,
${\vec{R}}_{\vec{\mu}} = (\hat{i}\mu_{1} + \hat{j}\mu_{2} + \hat{k}\mu_{3})a$, 
where $a$ is the lattice spacing, $\mu_{i}$ ($i=1,2,3$) are integers and 
$\hat{i}, \hat{j}$ and $\hat{k}$ are the unit vectors along the X, Y and Z 
axes, respectively, we make use of the well-known dispersion relation
for the tightly-bound energy bands \cite{callaway}
\begin{equation}
\epsilon_{n}(k) = \lambda_{n}(0) + 2\lambda_{n}(1)[  \cos k_{x} a + 
\cos k_{y} a + \cos k_{z} a ] .
\end{equation}
Here 
$\lambda_{n}(0) = \int d^{3}r_{A} \chi_{n}^{*}(\vec{r}_{A}-
\vec{R}_{\vec{\mu}}) H_{0}^{A} \chi_{n}(\vec{r}_{A}-\vec{R}_{\vec{\mu}})$,
$\lambda_{n}(1) = \int d^3r_{A}
 \chi_{n}^{*}(\vec{r}_{A}-\vec{R}_{\vec{\mu}})
H_{0}^{A} \chi_{n}(\vec{r}_{A}-\vec{R}_{\vec{\mu 1}})$, and  
$\vec{R}_{\vec{\mu 1}}=\vec{R}_{\vec{\mu}} + \hat{u}a$, $\hat{u}$ being any 
one of the unit vectors $\hat{i}, \hat{j}$ and $\hat{k}$. In the harmonic 
approximation of the lattice potential, $\chi_{n}(r)$ is the $n$th
level 3D harmonic oscillator eigenfunction.
Under these conditions, the unperturbed Green function can be
expressed in the chosen basis as \cite{koster}
\begin{eqnarray}
G_{nm}(\vec{\mu},\vec{\mu}^{\prime})  & = & \langle nm; \vec{\mu}
| \frac{1}{E^{+} - H_{0}}  |nm; \vec{\mu}^{\prime}\rangle \nonumber
\\ & = & 
-i C \int_{0}^{\infty} dt \exp(iE^{\prime}
t)[J_{\mu_{1+}}(t)J_{\mu_{2+}}(t)J_{\mu_{3+}}(t)+
J_{\mu_{1-}}(t)J_{\mu_{2-}}(t)J_{\mu_{3-}}(t)],  \hspace{0.5cm}
E'<3  \nonumber \\
& = &
C \int_{0}^{\infty} dt \exp(-E^{\prime}
t)[I_{\mu_{1+}}(t)I_{\mu_{2+}}(t)I_{\mu_{3+}}(t)+
I_{\mu_{1-}}(t)I_{\mu_{2-}}(t)I_{\mu_{3-}}(t)],  \hspace{0.5cm}
E'\ge 3  
\end{eqnarray}
Here $\mu_{i\pm} =|\mu_{i} \pm \mu_{i}^{\prime}|$, $J_{n}$ and $I_{n}$
 are the
$n$th order  Bessel functions, $E^{\prime} =
\{E-\lambda_{n}(0)-\epsilon_{m}(0)\}/\lambda_{nm}(1)$ are the
normalized energies and $C =
2{\mathcal N}^{2}/(2\pi\lambda_{nm}(1))$, with
$\lambda_{nm}(1)=2\{\lambda_{n}(1) +\epsilon_{m}(1)\}$.

Let us take the circularly-polarized additional laser field to have  
direction cosines ($1/\sqrt{3}, 1/\sqrt{3}, 1/\sqrt{3}$) 
in a SC optical lattice formed by far-off resonance blue-detuned 
lasers \cite{mueller}. Then, for two atoms lying along either of the X, Y or 
Z axes, the interatomic potential in Eq. (1) becomes 
\begin{eqnarray}
V_{AB} = - \frac{8\pi\alpha^2 Ik^2}{3c} \frac{\cos(kr/\sqrt{3}) 
\cos(kr)}{r}. 
\end{eqnarray}
The crucial point is that for two atoms residing at the nearest lattice 
sites, this potential is attractive for $r\simeq\pi/k_L\ll\pi/k$. 
By contrast, if the two atoms lie along the  
direction ($ 0, \frac{1}{\sqrt{2}},\frac{1}{\sqrt{2}}$), the corresponding
potential 
\begin{equation}
V_{AB} = \frac{2\pi\alpha^2 I}{c} \cos(\sqrt{2}kr/\sqrt{3}) 
\left [-\frac{5}{3} \cos(kr) + \frac{k\sin(kr)}{r^2}
 + \frac{\cos(kr)}{r^{3}} \right ]
\end{equation}
is repulsive for atoms at adjacent sites in this limit.
For near-zone separations $r\simeq\pi/k_L\ll\pi/k$, Eq. (11) reduces to the
``gravity''-like $1/r$ attractive potential.
The competition of such a potential with 
the
van der Waals (VdW) short-range repulsion ($s$-wave scattering) is analyzed in \cite{duncan}.
However, these VdW effects are irrelevant for atoms tightly
bound at  lattice sites separated by 100 nm or more. We therefore can impose a cut-off 
separation $r_{c}$($>r_{0}$), where $r_{0}$ is of the order of $10^3$ 
Angstroms, for calculating the low-energy collisional dynamics 
of the two atoms. 

We shall illustrate the
effect for two bosonic Li atoms that are
in the lowest band ($n=m=0$) of a SC lattice in the absence of the
additional laser ($V_{AB}=0$). 
The Lorentzian change in the density of states [Eq. (6)] is associated with
resonance states formed (for an appropriate $V_{AB}$) within the energy range of the two unperturbed atoms 
$-3<E'<3$, that is, as long as the two atoms remain confined within the lowest 
band. In contrast, a genuine bound  state (a resonance with zero width 
$\Gamma_r = 0$ in Eq. (4)) is formed at an energy exceeding that of the 
two unperturbed atoms in the lowest band ($E'>3$), that is, when at least 
one atom populates a higher band in the presence of $V_{AB}$. 

From Fig. 1 we see that for the laser
propagation direction 
as in 
Eq. (11), the interatomic potential 
is attractive, 
only at
$\cos\theta=\sqrt{1/3}$, for
$r\simeq\pi/k_L\ll\pi/k$. 
Hence, only these atoms would contribute to
the signature of a resonance (Fig. 2 - inset): the drastic 
increase of the two-atom density of states, i.e., the appearance of
distinct peaks around  $E_r$
(yet bearing in mind that
$\Gamma_{\rm heat}$ is their dominant experimental linewidth). 
The $E_r$ and $E_b$
in Fig.2 are shown to scale roughly linearly with
laser intensity (at a fixed detuning $\delta$). 
For the $^6S_{1/2}\rightarrow ^6P_{3/2}$ (852.1 nm) transition in Cs,
the recoil energy $E_R\alt 10$ kHz, whence conditions (a), (b) are best satisfied by a
near-UV lattice, $k_L\gg k$, $F_\theta(kr)\gg 1$,
$f_{\rm LD}(\vec{k}\cdot\vec{r})\ll 1$ and an additional
laser with $S\sim 10^{-4}$, $I\alt 0.1$ W/cm$^2$.

The solution of the Schroedinger equation, $H\Psi=E\Psi$ for the
tightly-binding limit and in the presence of $V_{AB}$ can be
written as
\begin{equation}
\Psi = \sum_{nm} \sum_{\mu \nu} C_{nm}|nm;\mu\nu\rangle .
\end{equation}
This is a superposition of the two-atom Wannier functions (8)
with coefficients $C_{nm}$ obtainable from the Lippman-Schwinger
matrix equation \cite{callaway}
\begin{eqnarray}
\sum_{\mu^{\prime}\nu^{\prime}} \langle
nm;\mu\nu|(H_{0}-E)|nm;\mu'\nu'\rangle C_{nm}(R_{\mu},R_{\nu}) \nonumber \\
& + & \sum_{\mu^{\prime}\nu^{\prime}}
V_{nm;\mu\nu,\mu^{\prime}\nu^{\prime}}
C_{nm}(R_{\mu^{\prime}},R_{\nu^{\prime}}) = 0 ,
\end{eqnarray}
where $V_{nm;\mu\nu,\mu^{\prime}\nu^{\prime}} = \langle nm; \mu
\nu| V_{AB}|nm; \mu^{\prime} \nu^{\prime} \rangle$. Equation(14)
can be solved by using the unperturbed Green's function Eq.(10).
The solution is given by
\begin{equation}
C_{nm}(R_{\mu},R_{\nu}) =  C_{nm}^{(0)}(R_{\mu},R_{\nu}) +
\sum_{\mu'\nu',\mu''\nu''}G_{nm}(\mu \nu,\mu'\nu')
V_{nm;\mu'\nu',\mu^{\prime\prime}\nu^{\prime\prime}}C_{nm}
(R_{\mu^{\prime\prime}},R_{\nu^{\prime \prime}}) ,
\end{equation}
where $C_{nm}^{(0)}$ pertains to the uncoupled superposition of
two-atom Wannier functions satisfying the equation
\begin{eqnarray}
\sum_{\mu^{\prime}\nu^{\prime}} \langle
nm;\mu\nu|(H_{0}-E)|nm;\mu'\nu'\rangle
C_{nm}^{(0)}(R_{\mu'},R_{\nu'})=0 .
\end{eqnarray}

The formation of a quasibound or resonance state gives rise to {\it
quantum translational correlations} ({\it spatial entanglement}) of 
the
two atoms. This entanglement is seen to correspond to {\em strong 
deviations} of the
$C_{nm}$ in Eq.(15) from the uncoupled coefficients $C_{nm}^{(0)}$, at $E$
near a resonance. Its experimental manifestation 
would be the {\it non-separability} and
redistribution of the 
two-atom density, as compared to the non-entangled (unperturbed) density of 
atoms at adjacent sites (Fig. 3). 

We have concentrated here on the essential effects 
of laser-coupled atom pairs at adjacent sites of a moderately
populated lattice, but
at higher densities we can obtain {\it multi-atom} correlated 
arrays.
A {\it
mean-field analysis} reveals the possibility of a bosonic ``supersolid''---a 
self-organized periodic array of atoms, under the influence of dipole-dipole 
interactions induced by a {\it circularly-polarized plane-wave laser} 
({\em without} an optical lattice) \cite{note}.

To conclude, we have demonstrated the possibility of observing a
novel effect: the quasi-binding or binding of cold atoms, initially 
residing at adjacent sites of an optical lattice, by a plane-wave
{\em off-resonant}
laser. 
Their controllable spatial entanglement
may be used for matter-wave
teleportation \cite{opa01} and other quantum information applications.
Experimentally, such giant quasimolecules can be revealed as satellite 
lines (with energies depending on the binding potential depth) shifted relative to 
the Bragg scattering spectral lines of a probe laser,
analogously to excitonic lines in 
crystals.

This work has been supported by the German-Israeli Foundation.
One of us (BD) thanks S. Giovanazzi for many helpful discussions.

\begin{figure}
\psfig{file=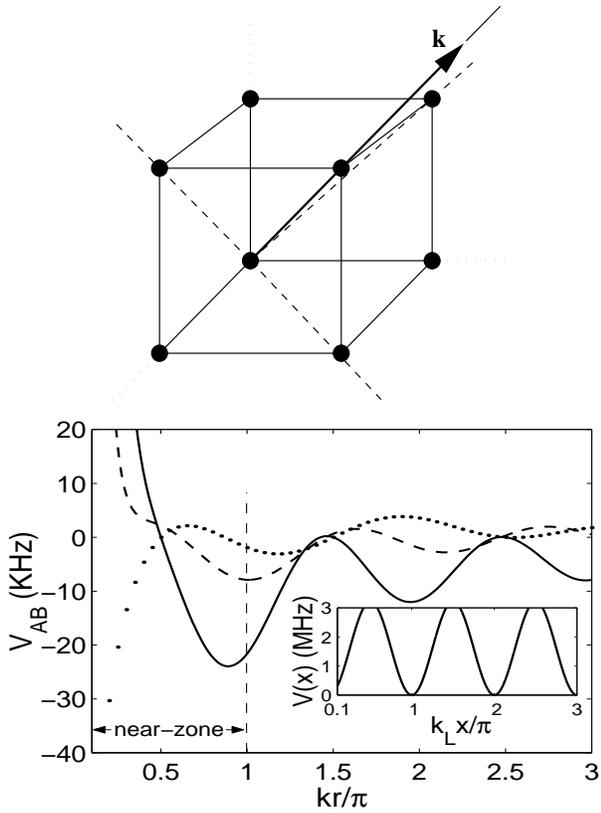,width=3.25in}
\caption{
The laser-induced off-resonant dipole-dipole potential for two Li
atoms in their ground states ($2S_{1/2}$) as a function of $kr$,
for  three 
interatomic axes at different  angles $\theta$ with the
 direction of propagation of the circularly polarized 
laser: $\cos \theta = 0 $ (solid
line), $\cos \theta = \sqrt{2/3} $ (dashed line), $\cos \theta =
\sqrt{1/3} $ (dotted line). The laser 
frequency is blue-detuned by 300
$\Gamma$ from the atomic transition $2S_{1/2} \to 2P_{3/2}$
(wavelength $ 670.77$ nm and linewidth $\Gamma/2\pi =
5.9$ MHz), $I=5$ W/cm$^2$, $S= 10^{-4}$. 
Inset (top) shows the geometry of the system. The other inset
shows the x-component of the periodic lattice potential 
$V_x = V_0 \sin ^2 k_L x$ 
generated by lasers 
blue-detuned by $10^4 \Gamma$ from  the same transition ($k\simeq
k_L$) with 
$S= 7.6 \times 10^{-5}$}. 
\end{figure}

\begin{figure}
\psfig{file=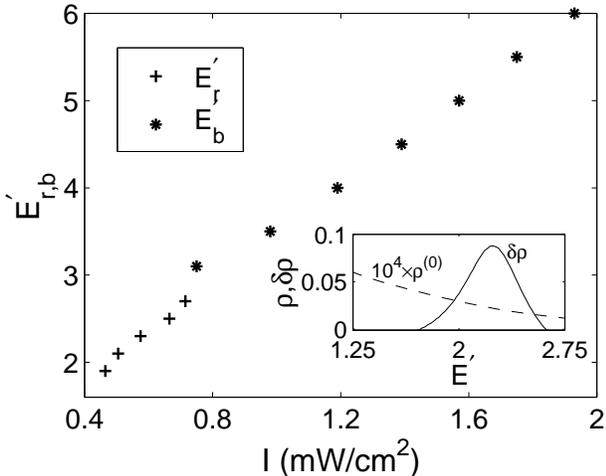,width=3.25in}
\caption{
Main figure: the dimensionless resonance energy 
$E'_r=\frac{E_r - 2 \lambda_0 (0)}{4 \lambda_0 (1)}$ (marked by ``+'') and
the bound state energy $E'_b=\frac{E - 2 \lambda_0 (0)}{4 \lambda_0 (1)}$
(marked by "*") 
as a function of laser intensity $I$ for the parameters of Fig. 1. 
Here $\lambda_0 (0) = 3.408$ MHz, 
$\lambda_0 (1)=0.98$ Hz. 
Inset:
the drastic increase of the two-atom density of states $\delta\rho$ 
by 4 orders of magnitude (solid line) 
compared to the unperturbed scaled density of states $\rho=10^{4}\times\rho^{(0)}$ (dashed line) 
near the resonance $E'_r=2$, for low laser intensity 
$I = 0.48$ mW/cm$^2$, $\delta = 300\Gamma$. The resonance 
linewidth 
$\Gamma_r=6.29$ Hz.} 
\end{figure}

\begin{figure}
\psfig{file=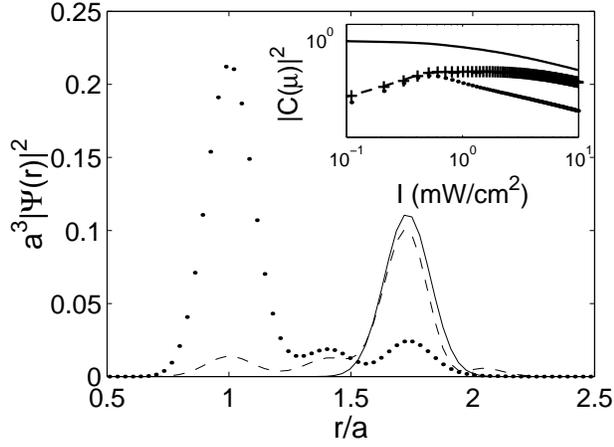,width=3.25in}
\caption{
Inset: Scattered intensity $|C(\vec{\mu})|^2$ as a
function of laser intensity $I$ ($\delta = 300\Gamma$). 
The incident atoms ($I=0$) are assumed to be in sites
$(000,111)$. Here solid line refers to $|C(000,111)|^2$, dashed
lines to $|C(000,100)|^2$, ``+'' to $|C(000,110)|^2$ and dotted line to
$|C(000,200)|^2$.
Main figure: Two-atom wavefunction squared (times the
unit cell volume $a^3$) 
$|\Psi(r)|^2=\int d^{3}\vec{R}|\Psi(\vec{R},\vec{r})|^2$,
where $\vec{R}$ 
is their center-of-mass coordinate,
as a function of
the separation $r$ (in units of $a$) for
$I = 0.48$ mW/cm$^2$ (dashed), $I = 5$ W/cm$^2$ (dotted) and $I = 0$ (solid).
All adjacent sites become partially occupied by the nonseparable
(entangled) atoms}.
\end{figure}

\end{document}